# HARDWARE IMPLEMENTATION OF (63, 51) BCH ENCODER AND DECODER FOR WBAN USING LFSR AND BMA

[1]Priya Mathew, [1]Lismi Augustine, [2]Sabarinath G., [2]Tomson Devis

[1,2]Department of ECE, St. Joseph's College of Engineering & Technology, Palai, Kerala
priyamathew.kpm@gmail.com, lismiaug@gmail.com
sabaripillai@ieee.org, tomsondevis@yahoo.com

## ABSTRACT

*Error Correcting Codes are required to have a reliable communication through a medium that has an unacceptable bit error rate and low signal to noise ratio. In IEEE 802.15.6 2.4GHz Wireless Body Area Network (WBAN), data gets corrupted during the transmission and reception due to noises and interferences. Ultra low power operation is crucial to prolong the life of implantable devices. Hence simple block codes like BCH (63, 51, 2) can be employed in the transceiver design of 802.15.6 Narrowband PHY. In this paper, implementation of BCH (63, 51, t = 2) Encoder and Decoder using VHDL is discussed. The incoming 51 bits are encoded into 63 bit code word using (63, 51) BCH encoder. It can detect and correct up to 2 random errors. The design of an encoder is implemented using Linear Feed Back Shift Register (LFSR) for polynomial division and the decoder design is based on syndrome calculator, inversion-less Berlekamp-Massey algorithm (BMA) and Chien search algorithm. Synthesis and simulation were carried out using Xilinx ISE 14.2 and ModelSim 10.1c. The codes are implemented over Virtex 4 FPGA device and tested on DN8000K10PCIE Logic Emulation Board. To the best of our knowledge, it is the first time an implementation of (63, 51) BCH encoder and decoder carried out.*

## 1. INTRODUCTION

Claude Shannon proposed the theorem of Channel capacity stating that, "Channel capacity is the maximum rate at which bits can be sent over the channel with arbitrarily good reliability"[1]..According to Channel Coding theorem,"The error rate of data transmitted over a band-limited noisy channel can be reduced to an arbitrarily small amount if the information rate is lower than the channel capacity" [2]. Error correcting codes are used in satellite communication, cellular telephone networks, body area networks and in most of the digital applications. There are different types of error correcting codes based on the type of error expected, expected error rate of the communication medium, and whether re-transmission is possible or not. Few of them are BCH, Turbo, Reed Solomon, Hamming and LDPC. These codes differ from each other in their implementation and complexity.

In 1960s Bose, Ray – Chaudhuri, Hocquenghem, independently invented BCH codes [3]. They are powerful class of cyclic codes with multiple error correcting capability and well defined mathematical properties. The Galois Field or Finite Field Theory defines the mathematical properties of BCH codes.

With decreasing size and increasing capability of electronic device, small and portable devices would be developed for communication around human bodies. Wireless body area network replace conventional healthcare systems by continuous health monitoring of patients. From implementation perspective, the required power levels of battery operated device limits the number of computational operations and power consumption of radio front-end. As radio front-end is main power consuming part, it determines the power consumption of a PHY and MAC of WBAN [4].

Simple block codes are preferred in the cases where low power consumption is required. In this work (63, 51) systematic BCH encoder and decoder is implemented. At the transmitter side the data streams are encoded by appending some extra bits with message bits known as parity bits. The message bits together with parity bits known as a 'Codeword'. At the receiver end codeword is checked for any error and then the corrupted bits are corrected. This is the decoding process. After correcting the errors in received data within the correction limits, the original message is retrieved. Thus improves the quality of transmitted message to a great extent and hence reduction in bit error rate.

The rest of this paper is organized as follows. Related works in section 2. WBAN is introduced in section 3 and BCH codes in section 4. BCH (63, 51) Encoder and Decoder are discussed section 5 and 6 respectively. In section 7 simulation results are discussed. Conclusions are drawn in section 8.

## 2. RELATED WORK

In [5] binary BCH codes for 5 bit correction in a 127 length bits using Peterson decoding procedure was discussed. Author shows that the burst error correcting codes are more complex in hardware. (7, 4) BCH encoder for encoding of a character and embedding in a binary document image was discussed in [6]. They can correct one bit error in any position of 7 bit data. The synthesis and simulations were carried out using Xilinx and implemented on Spartan 3 FPGA a (15, k) BCH encoder [7]. The result presented from the synthesis and timing simulation, shows the (15, 5, 3) BCH Encoder is more advantageous over the other two, according to speed requirement.

Proposed work discusses, hardware implementation of (63, 51) BCH Encoder and Decoder for WBAN using VHDL. This work aims to correct up to 2 errors in any position of 63 bit codeword. Here 51 bits are encoded into a 63 bit codeword. The decoder is implemented using the BMA and Chien search algorithm. Simulation was carried out using Xilinx 14.2 ISE simulator. Synthesis was carried out using the XST compiler and was successfully implemented on Virtex 4 FPGA.

## 3. WIRELESS BODY AREA NETWORK

The IEEE 802.15.6 is a standard for short-range, low power wireless communications in and around a human body [8]. Wireless Body Area Network (WBAN) offers many promising new applications in the area of remote health monitoring, home/health care, medicine, multimedia, sports and many others. The 15.6 standard defines a new communication standard with Medium Access Control (MAC) layer and three Physical (PHY) layers, namely Narrowband (NB), Ultra-wideband (UWB), and Human Body Communications (HBC) layers [9]. Among the PHY layers, narrowband 2.4 GHz PHY is the most mature. BCH (63, 51, 2) is an excellent double error correcting code that can be used in the baseband processing of NB PHY.

The Physical-layer service data unit (PSDU) transmission and reception is one of the major responsibilities of NB PHY. For the transmission, the PSDU should be transformed into PPDU. PPDU is created by pre-appending physical-layer preamble and physical-layer header to PSDU. During the reception, this physical-layer preamble and physical-layer header helps in the successful demodulation, decoding and delivery of the PSDU. As the physical layer parameters are needed to decode the PSDU properly the PHY header should be recover with zero errors. To increase the robustness of PHY header, baseband transmitter incorporates (31, 19) BCH encoder, which is a shortened code derived from a BCH (63, 51) code which can correct up to 2 errors. The PSDU is also encoded with a systematic (63, 51) BCH encoder.

# 4. BCH CODES

BCH codes forms a class of random multiple error-correcting cyclic codes. For any positive integer $m \geq 3$ and $t < 2^m - 1$, there exists a binary BCH code with the following parameters [15]:

Block length: $n = 2^m - 1$

Number of parity-check digits: $n - k \leq mt$

Minimum distance: $d_{min} \geq 2t + 1$.

BCH codes are subset of the Block codes. In block codes, the redundancy bits are added to the original message bits and the resultant longer information bits called "codeword" for error correction are transmitted. The block codes are implemented as (n, k) codes where n indicates the codeword and k the original information bits.

The generator polynomial g(x) of the t-error-correcting BCH code of length $2^m$ -1 is the lowest-degree polynomial over GF(2) which has $\alpha, \alpha^1, \alpha^2, \ldots, \alpha^{2t}$ as its roots. Let $\Phi(X)$ be the minimal polynomial of $\alpha i$. g(X) is the least common multiple (LCM) of $\Phi 1(X)$, $\Phi 2(X)$,.., $\Phi 2t(X)$.

Let $\alpha$ be the primitive element of $GF(2^6)$ for BCH (63, 51, t = 2) codes. Hence $1 + \alpha + \alpha^6 = 0$ is the primitive polynomial. The generator polynomial (GP) is given by [10].

$$g(x) = LCM(\phi_1 1(X), \phi_2(X))$$

$$g(x) = (1 + x + x^6)(1 + x + x^2 + x^4 + x^6)$$

$$g(x) = (1 + x^3 + x^4 + x^5 + x^9 + x^{10} + x^{12})$$

## 4.1. Galois Field

Error correcting codes operate over a large extent on powerful algebraic structures called finite fields. A finite field is often known as Galois field after Pierre Galois, the French mathematician [10]. A field is one in which addition, subtraction, multiplication and division can be performed on the field elements and thereby obtaining another element within the set itself. A finite field always contains a finite number of elements and it must be a prime power, say $q = p^r$, where p is prime. There exists a field of order q for each prime power $q = p^r$ and it is unique [11]. In Galois field GF(q), the elements can take this q different values.

We are exploiting the following properties of a finite field:

1. Addition and multiplication operations are defined.
2. The result of addition or multiplication of two elements is always an element in the field.
3. Zero is an element in the field, such that $a + 0 = a$ for any element a in the field.
4. Unity is an element in the field, such that $a \bullet 1 = a$ for any element a in the field.

For BCH (63, 51, t = 2), the field will be over $GF(2^6) = GF(64)$. If $\alpha$ is a primitive element in $GF(2^6)$ and p(x) a primitive polynomial, then the field elements of $GF(2^6)$ can be generated by using

$$p(x) = (1 + \alpha + \alpha^6)$$

## 5. BCH Encoder

The codewords are formed by adding the remainder after division of message polynomial with generator polynomial. All codewords are the multiples of generator polynomial. At the encoding side, the generator polynomials are not usually split as it will demand more hardware and control circuitry. The polynomial is used as such for encoding. The generator polynomial for BCH (63, 5, t = 2) is given by

$$g(x) = (1 + x^3 + x^4 + x^5 + x^9 + x^{10} + x^{12})$$

The parity bits are obtained by computing the remainder polynomial r(x) as:

$$r(x) = \sum_{i=0}^{11} r_i x^i = x^{12} m(x) \bmod g(x)$$

where m(x) is the message polynomial.

$$m(x) = \sum_{i=0}^{50} m_i x^i$$

$r_i$ = 0,…., 11 and $m_i$ = 0,….,50 are elements of GF(64). $m_{50}$ is the first message bit to be transmitted and $m_0$ is the last message bit, may be a shortened bit. The parity bits follow the order as follows: r11 first, followed by r10 and so on.

BCH codes are implemented as systematic cyclic codes. Hence can be easily implemented and the logic which implements encoder and decoder is controlled into shift register circuits [12]. The remainder r(x) can be calculated in the (n-k) linear stage shift register with the feedback connection at the coefficient of generator polynomial. (63, 51) BCH codeword are encoded as follows:

During the 1st clock cycle $m_{50}$ is given as input to the shift register. LFSR is initialized with seed value 0. From 1 to k clock cycles all the 51 messages bits are transmitted and the linear feedback shift register calculates the parity bits. The parity bits generated in the linear feedback shift register are taken in parallel. Thus the 12 parity bits are appended to the message bits. The serial in parallel out BCH encoder architecture is shown in the Figure 1.

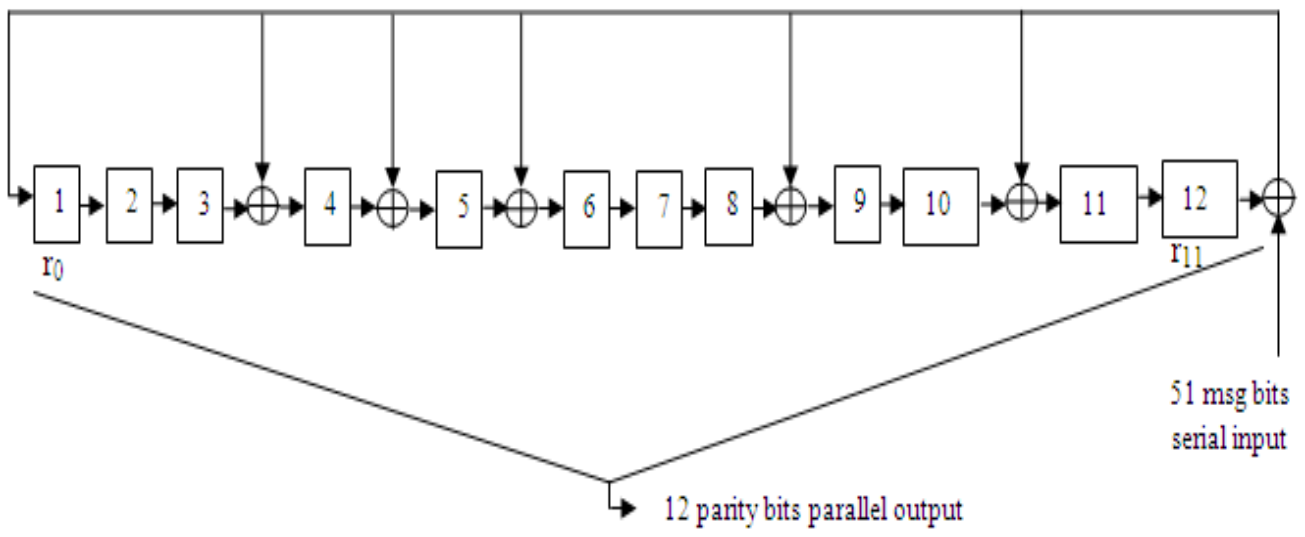

Figure 1. Block diagram of BCH (63, 51, 2) Encoder

## 6. BCH Decoder

Determining where the errors are in a received codeword is a rather complicated process. Figure 2 shows the block diagram of BCH (63, 51, 2) decoder. The decoding process for BCH codes consists of four major steps [13]:

1. Syndrome computation
2. Determine coefficients of error locator polynomial
3. Find the roots of error locator polynomial and it will indicate the erronous bits in the received codeword.

4. Error correction

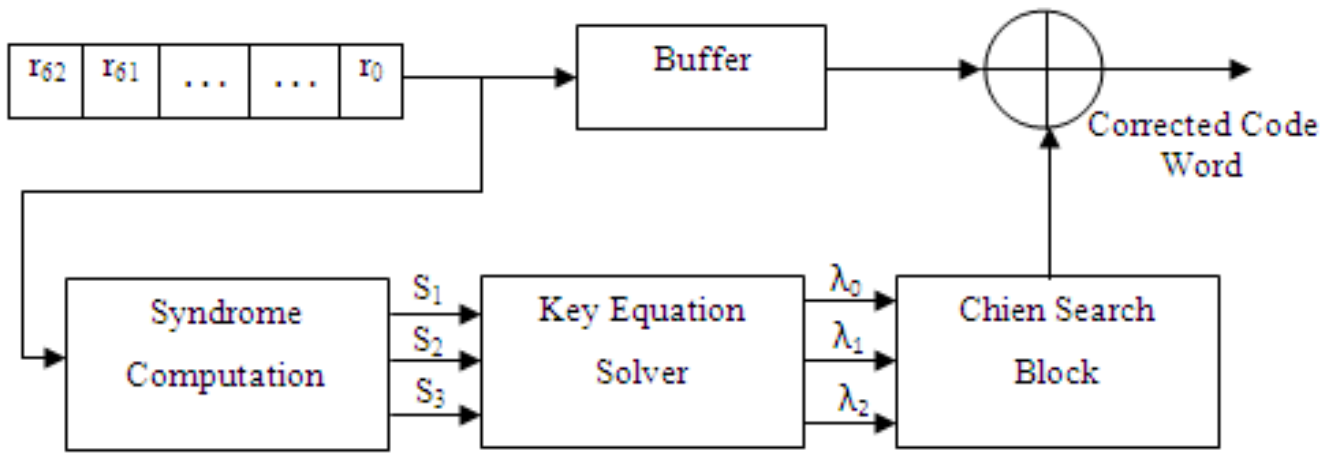

Figure 2. Block diagram of BCH (63, 51, 2) Decoder

## 6.1. Syndrome Computation

The first step at the decoding process is to store the received codeword in a buffer and then calculate the syndromes. The input to the syndrome module is the received codeword. The received polynomial may be corrupted with error pattern e(x) as shown [14].

$$r(x) = c(x) + e(x)$$

Where the received codeword is

$$r(x) = r_0 + r_1 x + .... + r_{n-1} x^{n-1}$$

Transmitted codeword is given by

$$c(x) = c_0 + c_1 + .... + c_{n-1} x^{n-1}$$

The error pattern is

$$e(x) = e_0 + e_1 + .... + e_{n-1} x^{n-1}$$

Syndrome $S_i$ can be computed by :

$$S_i = r(\alpha^i) = r_0 + r_1\alpha^i + r_2\alpha^{2i} + r_3\alpha^{3i} + .... + r_{n-1}\alpha^{(n-1)i}$$

where $1 \leq i \leq 2t - 1$.

For BCH (63, 51, t = 2) the three syndromes are:

$$S_1 = r(\alpha^1) = r_0 + r_1\alpha^1 + r_2\alpha^2 + r_3\alpha^3 + .... + r_{n-1}\alpha^{62}$$

$$S_2 = r(\alpha^2) = r_0 + r_1\alpha^2 + r_2\alpha^4 + r_3\alpha^6 + .... + r_{n-1}\alpha^{124}$$

$$S_3 = r(\alpha^2) = r_0 + r_1\alpha^3 + r_2\alpha^6 + r_3\alpha^9 + .... + r_{n-1}\alpha^{186}$$

where α is the primitive element in GF(26)

If there is no error in the received codeword then syndromes generated will be zero. Since the syndromes only depends on the error polynomial, and if the syndromes are nonzero then next step is to find out the coefficients of error locator polynomial.

## 6.2. Error Locator Polynomial Coefficients

The error locator polynomial is defined as:

$$\lambda(\alpha^j) = \lambda_0 + \lambda_1\alpha^j + \lambda_2\alpha^{2j}$$

In this paper Inversion-less Berlekamp Massey Algorithm (BMA) [15] is used for finding the coefficients $\lambda_0$, $\lambda_1$ and $\lambda_2$.

$$\lambda_2 = S_3 + S_1 S_2$$

$$\lambda_2 = S_1 S_1$$

$$\lambda_0 = S_1$$

The hardware implementation of syndrome and BMA is shown in the Figure 3. The efficient implementation of Galois field polynomial multiplication is an important prerequisite in BCH decoding process. We are implementing a new method in which MSE (Most Significant element) approach is used for finding the partial products [16]. Let a(α) and b(α) be two poynomials in $GF(2^6)$:

$$a(\alpha) = a_5\alpha^5 + a_4\alpha^4 + a_3\alpha^3 + a_2\alpha^2 + a_1\alpha^1 + a_0\alpha^0$$

$$b(\alpha) = b_5\alpha^5 + b_4\alpha^4 + b_3\alpha^3 + b_2\alpha^2 + b_1\alpha^1 + b_0\alpha^0$$

$$y(\alpha) = a(\alpha) * b(\alpha) = y_5\alpha^5 + y_4\alpha^4 + y_3\alpha^3 + y_2\alpha^2 + y_1\alpha^1 + y_0\alpha^0$$

Using the primitive polynomial $1 + \alpha + \alpha^6$, the partial products will be as follows:

$$y_5 = a_5(b_5 + b_0) + a_4b_1 + a_3b_2 + a_2b_3 + a_1b_4 + a_0b_5$$

$$y_4 = a_5(b_4 + b_5) + a_4(b_5 + b_0) + a_3b_1 + a_2b_2 + a_1b_3 + a_0b_4$$

$$y_3 = a_5(b_3 + b_4) + a_4(b_5 + b_0) + a_3b_1 + a_2b_2 + a_1b_3 + a_0b_4$$

$$y_2 = a_5(b_2 + b_3) + a_4(b_3 + b_4) + a_3(b_4 + b_5) + a_2(b_5 + b_0) + a_1b_1 + a_0b_2$$

$$y_1 = a_5(b_1 + b_2) + a_4(b_2 + b_3) + a_3(b_3 + b_4) + a_2(b_4 + b_5) + a_1(b_5 + b_0) + a_0b_1$$

$$y_0 = a_5b_1 + a_4b_2 + a_3b_3 + a_2b_4 + a_1b_5 + a_0b_0$$

The proposed method provides a fast multiplication algorithm which uses lesser number of ‘and’ and ‘xor’ gates.

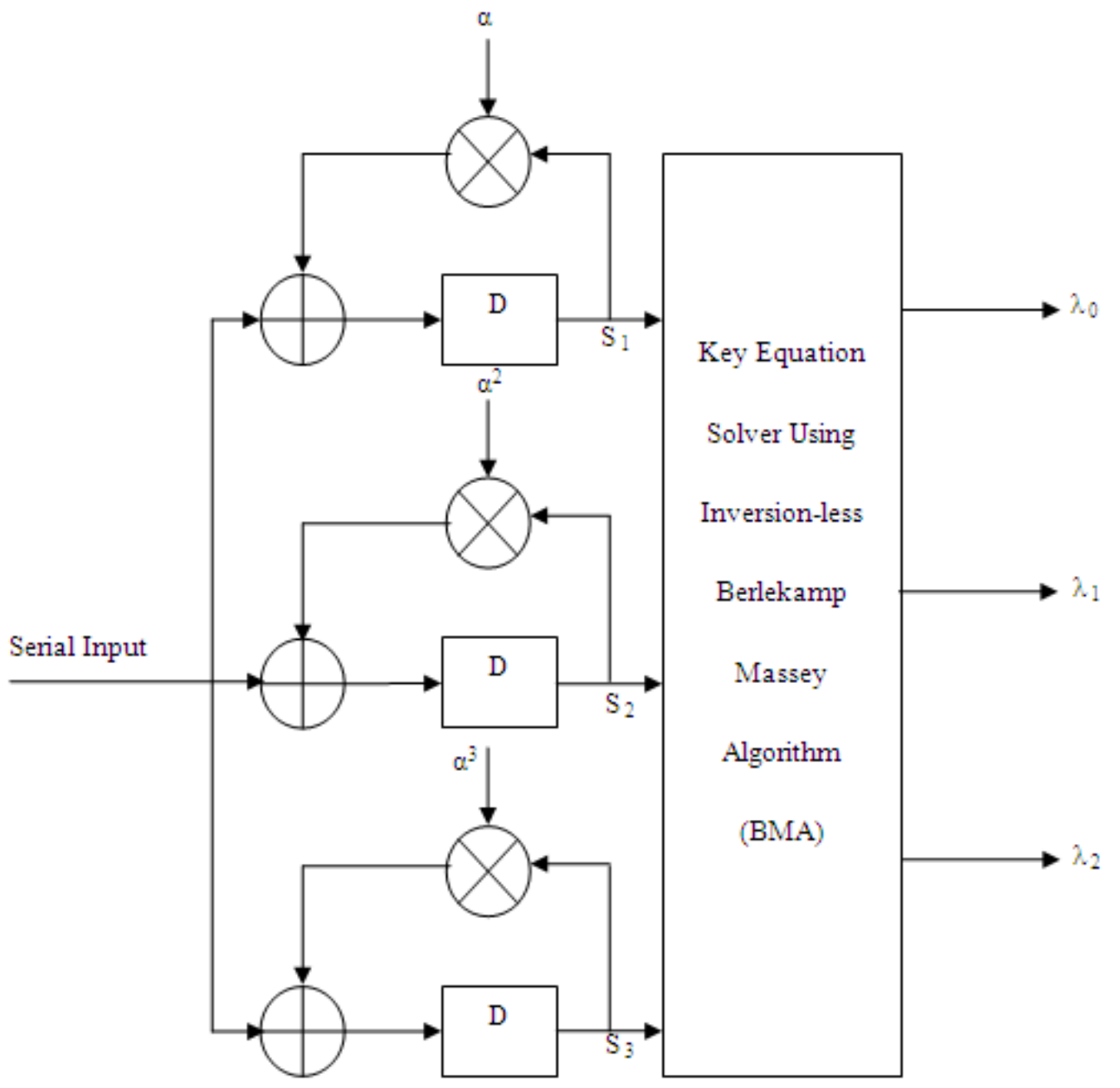

Figure 3. Implementation of Syndrome Calculator and Coefficient Solver

## 6.3. Finding Roots of Error Locator Polynomial

The next step in decoding process is to figure out the roots of the error locator polynomial. Chien search is an efficient algorithm for determining roots of polynomial defined over a finite field [17]. In $GF(2^6)$ all the 64 elements are examined to find whether it is a root or not. The hardware implementation of Chien search block is shown in Figure 4.

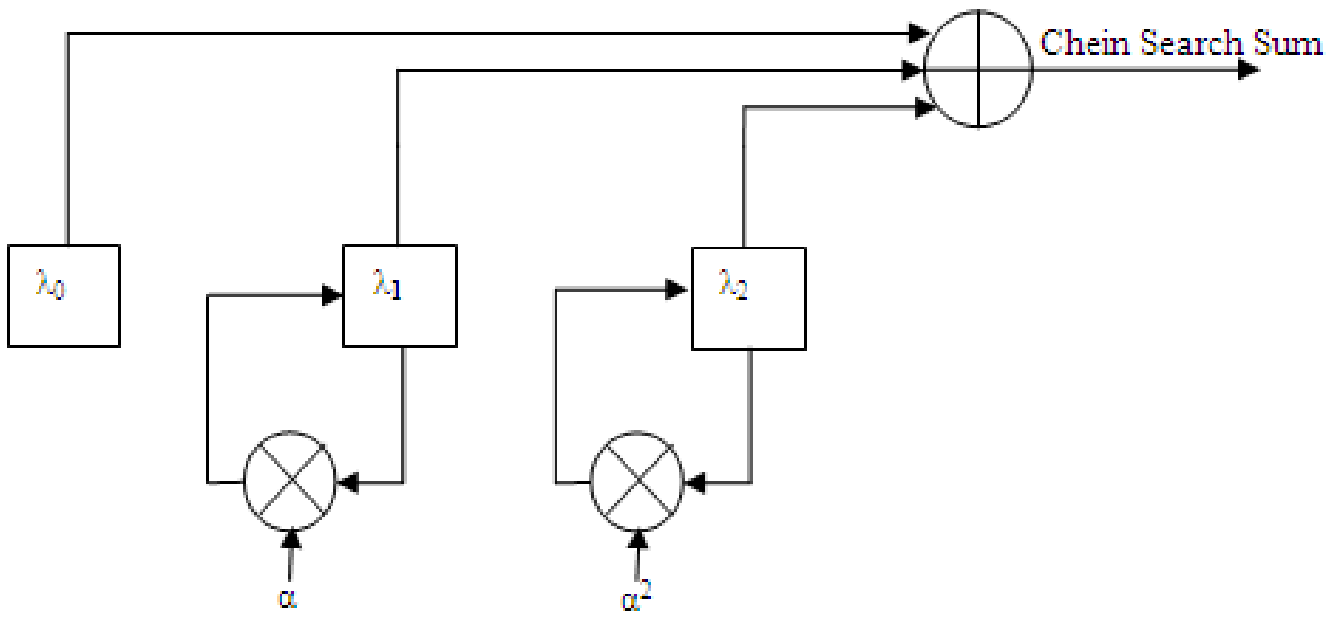

Figure 4. Implementation of Chien Search Block

The Chien search sum is given by:

$$\lambda_0 + \lambda_1 \alpha^{-j} + \lambda_2 \alpha^{2j}$$

where $1 \leq j \leq n$.

Let j be the clock cycle and if the Chien search sum equals to zero then $\alpha^j$ will be a root of the polynomial. Error location numbers are the reciprocal of the roots [18]. In order to find the error positions in the received codeword a counter is initialized with the Chien search. Therefore, if Chien search sum is zero it will indicate a root and magnitude of the root will be the counter value. Let $\alpha^j$ be a root, and then error position can be obtained by finding the inverse in the finite field.

$$\alpha^{-j} = \alpha^{-j} * \alpha^n$$

### 6.4. Error Correction

The final step in the decoding process is error correction. The received codeword is stored in a buffer register until it is corrected [19]. The erroneous bits can be corrected by simply flipping the bits at the error positions. Then output of the BCH decoder is the corrected codeword.

## 7. Results and Discussions

The proposed BCH encoder and decoder have been designed using VHSIC Hardware Description Language (VHDL) and simulated using ModelSim 10.1c. The results were also verified in MATLAB 7.8.0 Simulink model. In Matlab (63, 51) BCH Encoder Functional block has been modelled from the Communication Blockset and the input given as a frame based column vector. The design is synthesized using Xilinx ISE 14.2 and implemented on Xilinx XC4VLX200-FF1513 Virtex 4 FPGA. The results were tested and verified on DN8000K10PCIEe Logic Emulation board.

### 7.1. Simulation Results

Figure 5 and Figure 6 shows the simulation results of BCH encoder and decoder respectively. If the transmitted and the received codewords are the same then the syndromes will be zero. Here in this case the received codeword as erroneous is discussed. The received 63 bit encoded data given as input to the syndrome calculation circuit. Due to the presence of error the syndrome value will be a non- zero. Once the error is detected, it is corrected using BMA algorithm and Chien search algorithm as discussed.

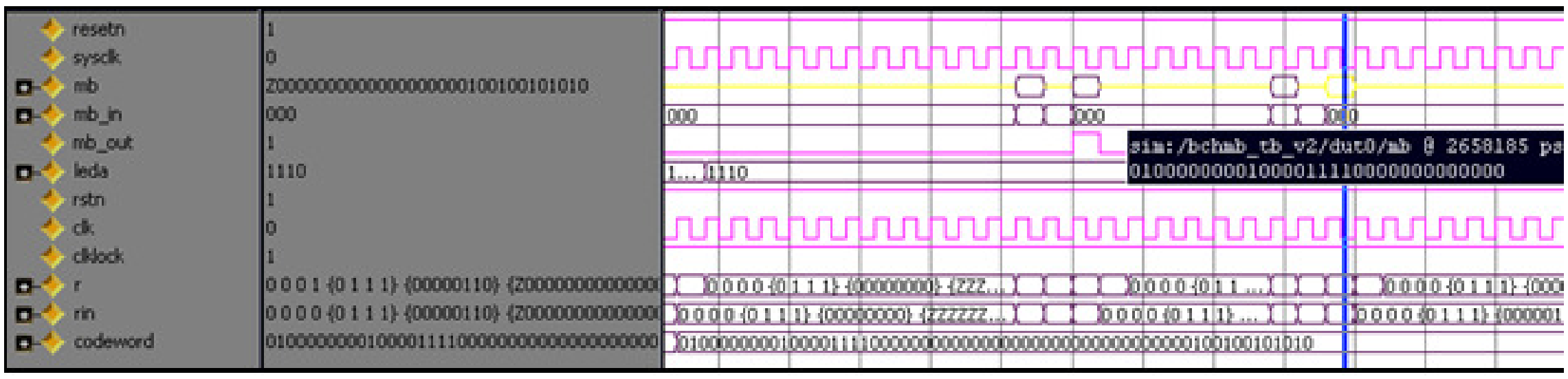

Figure 5. Simulation Results for (63, 51) BCH Encoder

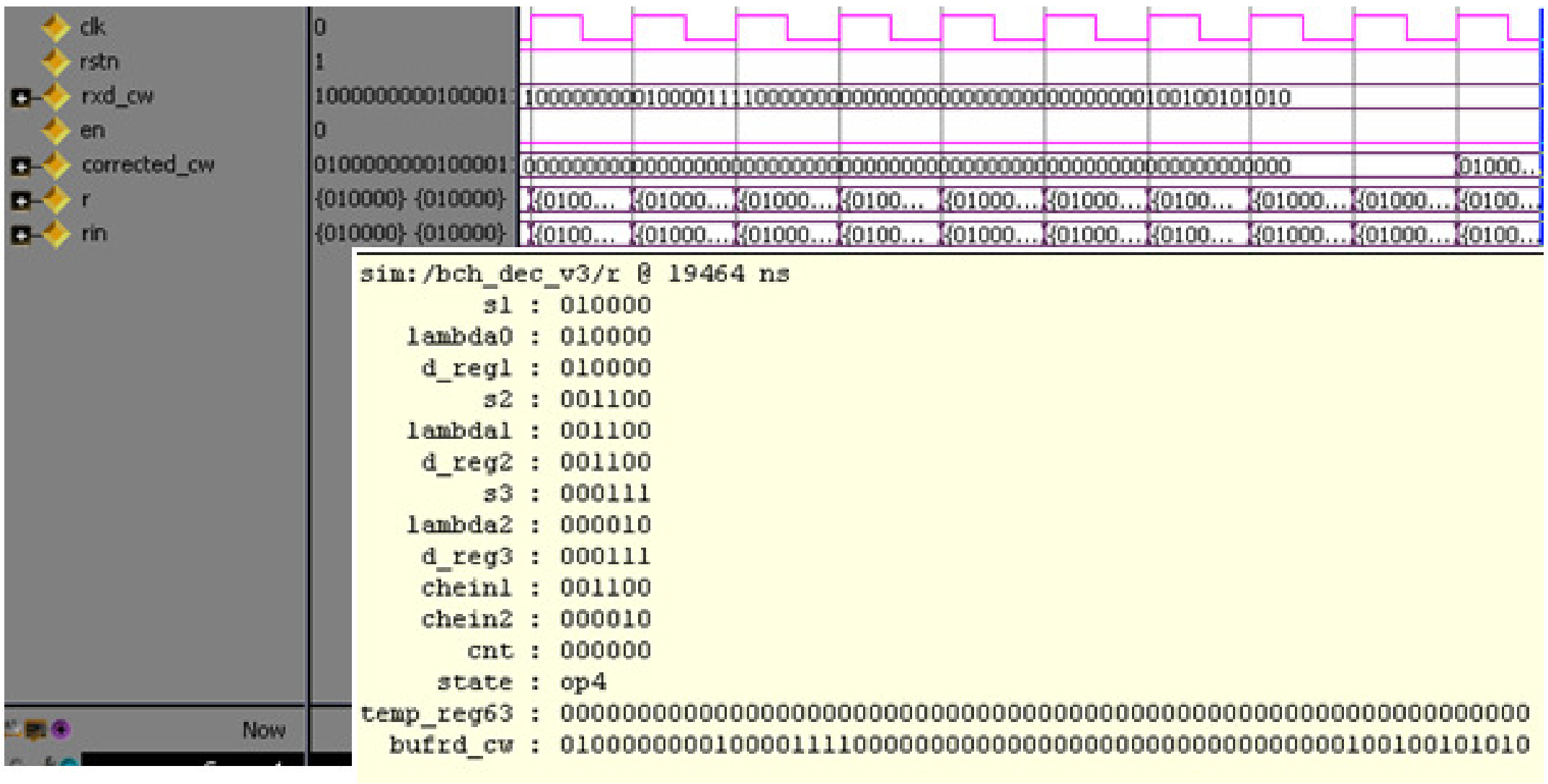

Figure 6. Simulation Results for (63, 51) BCH Decoder

## 7.2. Synthesis Results

Synthesis was carried out using Xilinx ISE 14.2 and tested in DN8000K10PCIE Evaluation Board. Figure 7 and Figure 8 shows the USB Controller Window of BCH encoder and decoder respectively. The device utilization summary of BCH encoder and decoder are given in Table 1 and Table 2 respectively.

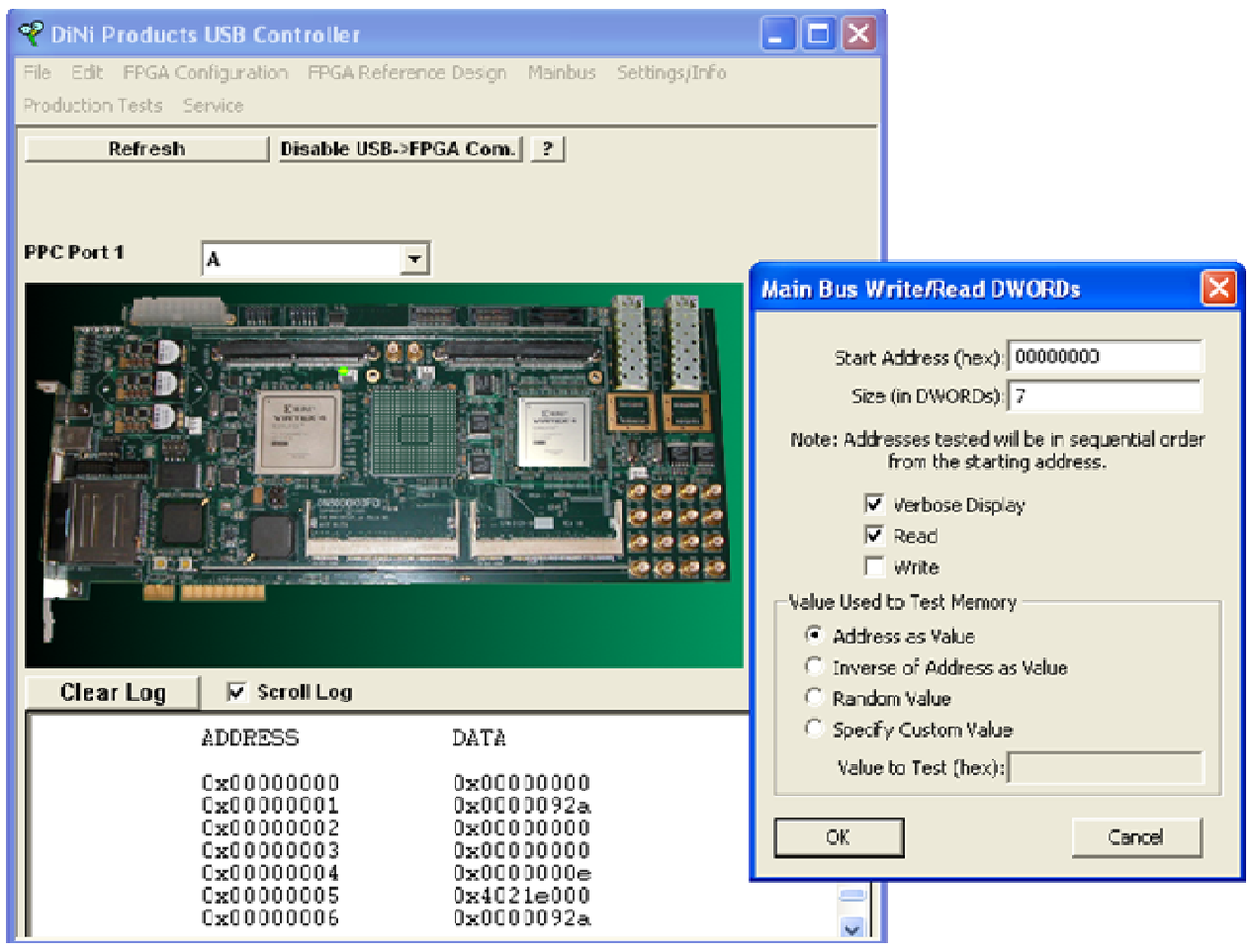

Figure 7. USB Controller Window showing BCH Encoder Output

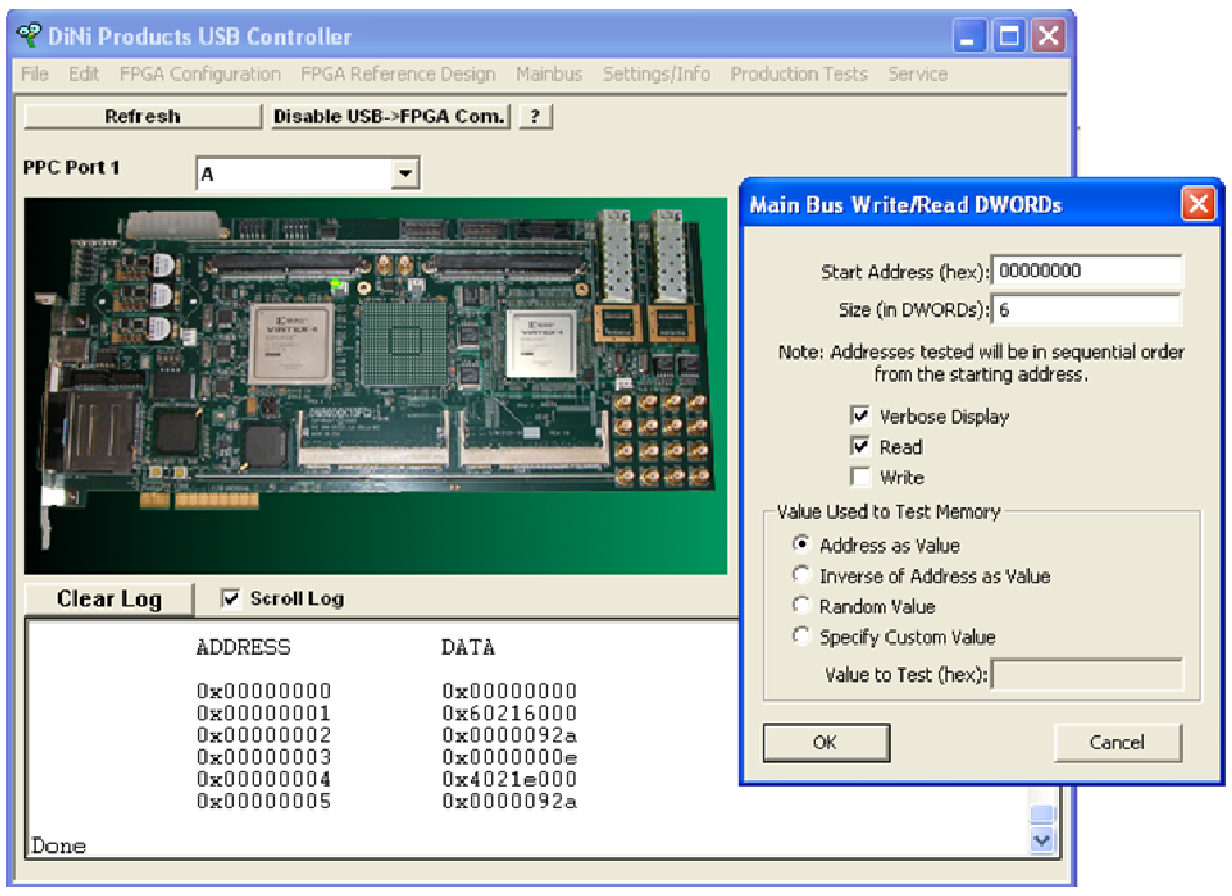

Figure 8. USB Controller Window Showing BCH Decoder Output

Table 1. Device Utilization Summary of BCH Encoder

| Device Utilization Summary | | | | [-] |
|---|---|---|---|---|
| **Logic Utilization** | **Used** | **Available** | **Utilization** | **Note(s)** |
| Number of Slice Flip Flops | 179 | 178,176 | 1% | |
| Number of 4 input LUTs | 389 | 178,176 | 1% | |
| Number of occupied Slices | 246 | 89,088 | 1% | |
| Number of Slices containing only related logic | 246 | 246 | 100% | |
| Number of Slices containing unrelated logic | 0 | 246 | 0% | |
| Total Number of 4 input LUTs | 389 | 178,176 | 1% | |
| Number of bonded IOBs | 42 | 960 | 4% | |
| IOB Flip Flops | 60 | | | |
| Number of BUFG/BUFGCTRLs | 2 | 32 | 6% | |
| Number used as BUFGs | 2 | | | |
| Number of DCM_ADVs | 1 | 12 | 8% | |
| Average Fanout of Non-Clock Nets | 3.06 | | | |

Table 2. Device Utilization Summary of BCH Decoder

| Device Utilization Summary | | | | [-] |
|---|---|---|---|---|
| **Logic Utilization** | **Used** | **Available** | **Utilization** | **Note(s)** |
| Number of Slice Flip Flops | 295 | 178,176 | 1% | |
| Number of 4 input LUTs | 818 | 178,176 | 1% | |
| Number of occupied Slices | 422 | 89,088 | 1% | |
| Number of Slices containing only related logic | 422 | 422 | 100% | |
| Number of Slices containing unrelated logic | 0 | 422 | 0% | |
| Total Number of 4 input LUTs | 819 | 178,176 | 1% | |
| Number used as logic | 818 | | | |
| Number used as a route-thru | 1 | | | |
| Number of bonded IOBs | 42 | 960 | 4% | |
| IOB Flip Flops | 71 | | | |
| Number of BUFG/BUFGCTRLs | 2 | 32 | 6% | |
| Number used as BUFGs | 2 | | | |
| Number of DCM_ADVs | 1 | 12 | 8% | |
| Average Fanout of Non-Clock Nets | 3.43 | | | |

## 8. CONCLUSION

To ensure reliable transmission of information through a physical medium or wireless medium, error control coding are used in the digital information and communication systems. In this paper we have presented the implementation of (63, 51, t = 2) BCH encoder and decoder which finds its application in IEEE 802.15.6 WBAN. Synthesis was successfully done using Xilinx ISE 14.2 and design implemented on XC4VLX200 Virtex 4 FPGA. Here 51 message bits are encoded into a 63 bit codeword. If there is any 2 bit error in any position of 63 bit codeword, it can be detected and corrected. The encoder is implemented using LFSR. The decoder uses Berlekamp algorithm and Chien Search algorithm. The proposed Galois field polynomial multiplication is used for the syndrome calculation as well as for finding the error locating polynomial coefficients. It allows fast field multiplication. BCH code forms a large class of powerful random error-correcting cyclic codes. They are relatively simple to encode and decode. Further, the performance can be improved by adopting parallel approach methods.

## REFERENCES

[1] Michelle Effros, Andrea Goldsmith, Yifan Liang, "Generalizing Capacity: New Definitions and Capacity Theorems for Composite Channels", IEEE Transactions on Information Theory, vol. 56, no. 7, pp. 3069-3087, July 2010.

[2] Wilfried Gappmair," Claude E. Shannon: The 50th Anniversary of Information Theory", IEEE Communications Magazine, pp. 102-105, April 1999.

[3] Shyue-Win Wei, Che-Ho Wei, "High-speed hardware decoder for double-error-correcting binary BCH codes", IEEE Proceedings, vol 136, no. 3, pp. 227-231, June 1989.

[4] Sana Ullah and Mohammed A. Alnuem, "A Review of IEEE 802.15.6 MAC, PHY, and Security Specifications", Hindawi Publishing Corporation International Journal of Distributed Sensor Networks, article ID 950704, pp:1-12, 2013.

[5] Laurencin Mihai Ionesco, Constantin Anton, "Hardware Implementation of BCH Error-Correcting Codes on a FPGA", International Journal of Intelligent Computing Research, vol 1, Issue 3, June 2010.

[6] P. Reviriego, C. Aggrades, J. A. Maestro, "Efficient error detection in Double Error Correction BCH codes for memory applications", Microelectronics Reliability, pp. 1528–1530,2012.

[7] Amit Kumar Panda, Shahbaz sarik , Abhishek Awasthi, "FPGA Implementation of Encoder for (15, k) Binary BCH Code Using VHDL and Performance Comparison for Multiple Error Correction Control", International Conference on Communication Systems and Network Technologies, pp. 780-784, 2012.

[8] IEEE Computer Society, "IEEE Standard for Local and metropolitan area networks: Part 15.6 Wireless Body Area Networks," IEEE Standards Association, 29, Feb., 2012.

[9] Kyung Sup Kwak, Sana Ullah, and Niamat Ullah, "An Overview of IEEE 802.15.6 Standard", IEEE Trans. 2012.

[10] R. Lidl and H. Niederreiter, "Finite Fields", Cambridge University Press, Cambridge, 1966.

[11] John Gill, "Finite Fields", Stanford University.

[12 D.Muthiah, A. Arockia Bazil Raj, "Implementation of High-Speed LFSR Design with Parallel Architectures", International Conference on Computing Communication and Applications (ICCCA), pp. 1-6, Feb. 2012.

[13] Rohith S, Pavithra S "FPGA Implementation of (15,7) BCH Encoder and Decoder for Text Message", International Journal of Research in Engineering and Technology,vol. 2, pp. 209-214,Sep 2013.

[14] Nur Ahmadi, M. Hasan Sirojuddin, A.. Dipta Nandaviri, Trio Adiono,"An Optimal architecture for BCH Decoder", International Conference on Application of Information and Communication Technologies (AICT), pp. 1- 5, 2010.

[15] James L. Massey, "Step-by-step Decoding of the Bose-Chaudhuri-Hocquenghem Codes", IEEE Transcations on Information Theory, vol. –IT 11, no. 4, pp 580-585.

[16] Miguel Morales-Sandoval, Arturo D´ıaz-Perez, "Compact FPGA-based hardware architectures for GF($2^m$) multipliers", 16th Euromicro Conference on Digital System Design, pp. 649-652, 2013.

[17] R.T Chien, "Cyclic Decoding Procedures for Bose-Chaudhuri-Hocquenghem Codes", IEEE Transcations on Information Theory, pp.357-363, Oct. 1964.

[18] Hardik Sutaria, Deepti Khurge, "FPGA based BCH Decoder", International Journal for Scientific Research & Development, vol. 1, pp. 637-640, 2013.

[19] R. Lidl and H. Niederreiter, "Finite Fields", Cambridge University Press, Cambridge, 1966.

[20] Bhasker J, "A VHDL Primer", P T R Prentice Hall, 1992.